\documentstyle[11pt,newpasp,twoside,epsf]{article}
\def\edcomment#1{\iffalse\marginpar{\raggedright\sl#1\/}\else\relax\fi}
\reversemarginpar

\begin{document}
\title{Early Disk Evolution}
\author{Mark Wardle}
\affil{Macquarie University, Sydney NSW 2109, Australia}

\begin{abstract}
A variety of processes play a role in the evolution of protostellar
disks.  Here I focus on the uncertain issue of magnetic field-disk
coupling and its implications for magnetically-driven turbulence and
disk-driven winds.  At present it is clear that the magnetic field
plays a crucial role in disk evolution, but detailed conclusions
cannot be drawn because the complicated interplay between dynamics and
the evolution of the grain population remains to be explored.
\end{abstract}

\section{Introduction}

Disks are the natural byproduct of the formation of stars: the resting
place of material with too high an angular momentum to participate
fully in the initial gravitational collapse that forms the central
protostar.  Accretion of an appreciable fraction of the stars mass may
(or may not) occur through the disk, and disks appear to be an
essential component of the engine that powers protostellar jets.  A
somewhat bewildering array of physical processes come into play to
determine disk structure and must be addressed if we wish to
understand disk evolution, forming planetary systems, and especially
to interpret essentially unresolved observations of disks.

To zeroth order the material in a disk is in orbit about the central
object: in other words, the radial component of the gravitational field
from the protostar is balanced primarily by rotation.  Even moderately
flattened structures supported by other means tend to be unstable to
interchange instabilities - whether the support is by gas pressure, radiation
pressure or magnetic stresses (e.g.\ Stehle \& Spruit 2001), and must
therefore be evolving rapidly.  This implies that the internal signal
speed in a disk that survives for many orbits is much less than the
Keplerian speed.   A slowly-evolving disk is thin
because its scale height is determined by the balance between vertical
pressure gradient and the tidal field of the central object.

One consequence is that the local dynamical time scale (i.e.\ the
orbital time scale) is much shorter than the time scale on which the
disk evolves.  This makes physical sense, as a well-defined disk
cannot otherwise be expected to exist for more than a few orbits.
There is however, a flip side to this: the range of orbital time
scales across the radial extent of the disk makes globally steady
evolution most unlikely.  It is unreasonable to expect that the
radial variation of the processes that determine disk structure
conspires to force the local behaviour to be just that needed to
guarantee that $\dot{M}$ is constant all radii.  Disk models that rely
on this assumption to tie together the behaviour at different radii
should be regarded with caution, though they \emph{might} represent
the disk in a time-averaged sense.

A variety of processes play a role in the evolution of protoplanetary
disks.  Some factors are well-defined and operate on a scale large
enough to be amenable to observation, such as tidal fields and
irradiation by UV from companion stars or nearby stars, the X-ray
irradiation from the central star, and self gravity if the disk mass
$\ga$30\% of the stellar mass.  Gravitational
disturbance by planets forming within the disk
is, of course, critical in the later stages of disk evolution.

The factors that operate on small-scales are not directly observable,
requiring a resolution equivalent to the disk thickness or less.  They
do however control transport processes, the internal dynamics of the
disk, and chemical evolution of the disk material.  (This latter point
may yield a way of teasing apart the processes that occur within
disks.)  Although much of the theoretical work on the physics
operating on these scales is directed towards the identification of
mechanisms responsible for angular momentum transport, note that
protoplanetary disks need not necessarily be \emph{accretion} disks.
Although magnetospheric accretion and emission from boundary layers
are observed in some systems, the total mass accreted onto the star
may be quite small compared to the total disk mass, and so accretion
may not be the defining characteristic of the disk structure.

Convection was an early proposed mechanism for the origin of angular
momentum transport in protostellar disks (e.g.\ Lin \& Papaloizou
1980).  The idea was that convective cells couple different disk
radii, thereby acting as an effective viscosity.  The convection would
be driven by internal heating due to the disipation associated wioth
the energy lost by the accreting material.  However, simulations show
that angular momentum is transported inwards rather than outwards
(e.g.  Cabot et al.\ 1987a,b), and in any case the convective cells
are thin in the radial direction (ironically because of angular
momentum conservation) and would be relatively inefficient as a
mechanism for driving accretion.

The leading contender is now the magnetic field, which can transport
angular momentum through the generation of turbulence via the
magnetorotational instability (Balbus \& Hawley 1991,2000; Balbus
2003), or through the centrifugal acceleration of material from the
disk surfaces and the formation of a jet (e.g. Blandford \& Payne 1982;
K\"onigl \& Pudritz 2000).  At present, it is not known what the
initial radial distribution of mass and magnetic flux is.  However, the
initial poloidal field will have a strength of at least several
milligauss, even if the ambient molecular cloud field is not
particularly dragged in during the collapse that forms the disk.  Even
this weak field is sufficient to act as a seed for the
magnetorotational instability at 1 AU (Salmeron \& Wardle, in
preparation).

Protostellar disks are weakly ionized, so the role
of the magnetic field is complicated by its dependence on the trace
charged species in the gas.  Thus the evolution of protostellar disks
is linked to chemistry, which determines the thermal and ionisation
structure of the disk, and also to the population of dust grains which
evolves through the accumulation of ices, sticking or shattering as a
result of grain collisions, advection by turbulence, grain drag
effects, and settling to the midplane (e.g.\ Weidenschilling \& Cuzzi
1993).

\section{Magnetic Coupling}

The ability of the magnetic field to drive turbulence via the MRI or
to produce a disk-driven wind is determined by its coupling to the
weakly-ionised fluid.  This is achieved by the collisions of charged
species drifting in response to the electric and magnetic fields.  The
coupling is sensitive to ionisation rate, density and the very
different mobilities of electrons, ions and grains.

The magnetic field in a weakly-ionised medium evolves according to
the induction equation
\begin{equation}
    \frac{\partial\mathbf{B}}{\partial t} = \mathbf{\nabla\times(v\times
    B) - c\nabla\times E'}
    \label{eq:induction}
\end{equation}
where $\mathbf{E'}$ is the electric field in the rest frame of the
fluid, which is related to the
current density $\mathbf{J} = c/4\pi \,\, \mathbf{\nabla\times B}$ by
\begin{equation}
\mathbf{E'} = \frac{\mathbf{J}}{\sigma_{\parallel}} +
    \frac{\sigma_{\mathrm{H}}}{\sigma_{\mathrm{H}}^2 +
           \sigma_{\mathrm{P}}^2} \frac{\mathbf{J\times B}}{B} -
    \left(\frac{\sigma_{\mathrm{P}}}{\sigma_{\mathrm{H}}^2 +
\sigma_{\mathrm{P}}^2} -
    \frac{1}{\sigma_\parallel}\right) \frac{(J\times B)\times
    B}{B^2} \,.
    \label{eq:E}
\end{equation}
Here $\sigma_\parallel$, $\sigma_{\mathrm{H}}$ and
$\sigma_{\mathrm{P}}$ are the field-parallel, Hall and Pedersen
conductivities and the three terms involving $\mathbf{J}$,
$\mathbf{J\times B}$ and $(\mathbf{J\times B})\mathbf{\times B}$
correspond to Ohmic, Hall, and ambipolar diffusion respectively.

Generally there is a trend from ambipolar to Hall and then Ohmic
diffusion with increasing density and decreasing ionisation rate,
which is why calculations of core collapse and disk formation have
generally assumed that ambipolar diffusion dominates whereas solar
nebula studies traditionally employ a resistive approximation.  However,
the Hall regime spans a surprisingly wide range of intermediate
conditions and cannot be neglected.  Furthermore, Hall diffusion
qualitatively changes the vector evolution of the magnetic field,
exhibiting a dependence on the strength of the magnetic field and a
sensitivity to handedness that breaks many simple planar geometries
that could otherwise be maintained in the presence of ambipolar or
Ohmic diffusion.

The broad range of ionisation rates and densities obtaining over the
radial and vertical extent of protostellar disks means that each form
of diffusion will dominate in different regions.  The ionisation rate
is dominated by x-rays from the central low-mass star in the upper few
g\,cm$^{-2}$ of the disk (Glassgold, Najita \& Igea 1997) and then by
cosmic rays to depths of a few hundred g\,cm$^{-2}$ (Hayashi 1981).
The increase in ionisation rate and decrease in density with
increasing height produces a trend from Ohmic through Hall to
ambipolar diffusion with increasing height.  If sufficient electrons
are present Ohmic diffusion may never become completely dominant even
at the midplane.  The biggest uncertainty in estimates of the degree
of coupling of the magnetic field to the disk is the grain population,
as ions and electrons stick to grain surfaces and the mobility of the
typical charge carriers is much reduced.  If, on the other hand,
grains have settled to the disk midplane, Hall diffusion dominates at
the midplane over disk radii from 0.1 to 100 AU (Sano \& Stone 2002a).

\begin{figure}[t]
\centerline{\epsfxsize=7.8cm \epsfbox{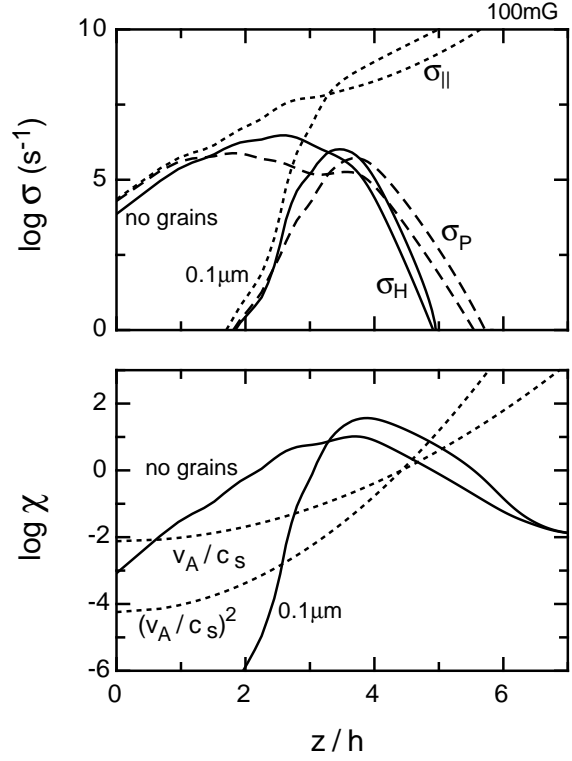}}
\caption{ \emph{Upper panel:} conductivity tensor components as a
function of height above the midplane at 1AU in a minimum solar nebula
for a dusty disk with 0.1$\micron$ grains and a model in which grains
have settled to the midplane.  The magnetic field strength is assumed
to be 100 mG. \emph{Lower panel:} The vertical profile of a coupling
parameter $\chi$ in the two models (solid lines), and the ratio of the
Alfven speed to the sound speed and the square of that ratio (dotted
lines).  The magnetic field effectively couples to the disk material
when $\chi\la v_{A}/c_{s}$; this criterion becomes $\chi\la
v_A^2/c_s^2$ in the presence of significant Hall diffusion.  }
\end{figure}

By way of example, the resulting vertical profile of the conductivity
tensor in a minimum solar nebular model at 1AU is plotted in the
presence and absence of 0.1 $\micron$ grains in the top panel of Fig.\
1.  The much larger conductivities
in the no-grain model within three scale heights of the midplane
reflects the lack of grains -- the charge is carried by more mobile
free ions and electrons.  The decoupling of ions from the magnetic
field for densities $\ga 10^{9}$\,cm$^{-3}$ means that Hall component
is larger than the Pedersen component beween 1.5 and 4 scale heights,
and is 80\% of $\sigma_{\mathrm{P}}$ even at the midplane.  In the 0.1
$\micron$ grain model, the Hall conductivity dominates the Pedersen
conductivity within 4 scale heights of the midplane.

The variation of magnetic coupling with height above the midplane
determines how deeply magnetic activity penetrates the disk and
therefore the thickness of the ``dead zone'' (Gammie 1996) which is not
affected by magnetic activity, and does not participate in accretion.

\section{Magnetorotational Instability}

One measure of whether the conductivity is sufficient for the magnetic
field to interact with the disk material is whether the field is
unstable to the magnetorotational instability.  This is determined by
comparing the coupling parameter $\chi = \omega_c /\Omega$ to the ratio
of Alfv\'en speed to sound speed, $v_A/c_s$.  Here $\omega_c$ is the
frequency above which ideal MHD breaks down and $\Omega$ is the
Keplerian frequency.  If Hall diffusion is unimportant the criterion
is $\chi\ga v_A/c_s$, if it is dominant (and the magnetic field has
the correct orientation), then $\chi\ga (v_A/c_s)^2$ is necessary.

The coupling parameter is plotted as a function of height in the lower
panel of Fig.\ 1.  The entire thickness of the disk is magnetically
active in the absence of grains, whereas in the single-size grain
model the layers above 2.5 scale heights are active.  In both cases,
Hall diffusion is important throughout the active regions.

If grains have settled to the disk midplane, Hall diffusion dominates
at the midplane over disk radii from 0.1 to 100 AU (Sano \& Stone
2002a).  The implications for the magnetorotational instability are
twofold.  First, in the presence of Hall diffusion the instability
operates under much weaker coupling than in the resistive or ambipolar
diffusion cases, and therefore the potentially magnetically turbulent
region is much more extensive than previously thought (Wardle 1999):
the magnetic ``dead zone'' in which the field does not couple to the
disk material is more restricted.  This is illustrated in Fig.\ 2,
which compares the linear development of the instability in models
with and without Hall diffusion (Salmeron \& Wardle 2003).  If Hall
diffusion is neglected, the instability grows in the upper layers of
the disk and there is a substantial dead zone ($z/h\la 2$).  With Hall
diffusion, the growth rate of the most unstable mode is increased
somewhat and the mode penetrates deeper into the disk, to within a
scale height of the midplane.
\begin{figure}[t]
\centerline{\epsfxsize=7cm \epsfbox{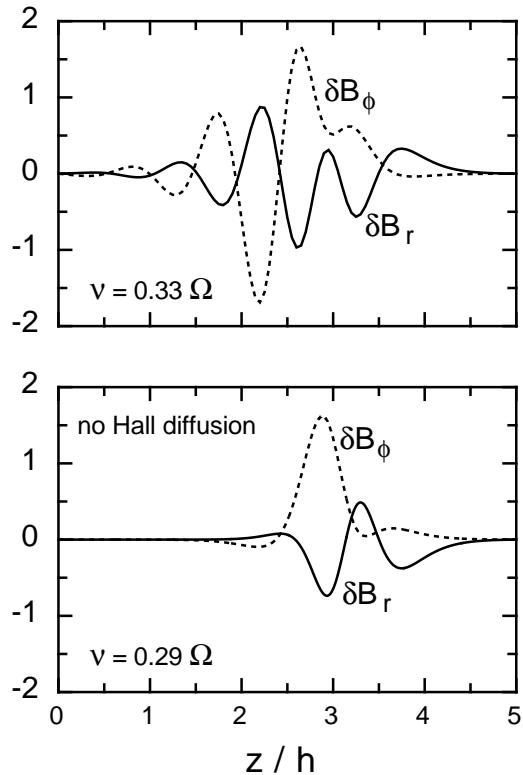}}\vskip 0cm
\caption
{Comparison of the structure and growth rate of the fastest growing
mode of the magnetorotational instability in a uniform vertical field
for $v_A/c_s = 0.01$.  Top panel shows the case where Hall diffusion is
dominant close to the midplane and ambipolar diffusion dominates near
the surface.  Bottom panel show the fastest growing mode when Hall
diffusion is neglected.}
\end{figure}

Second, in the nonlinear regime the turbulence grows and decays cyclically: as
the magnetic field is amplified through the nonlinear growth of the
instability, the charged particles become more tightly coupled to the
magnetic field, Hall diffusion becomes less important and growth shuts
off.  The field then decays until Hall diffusion becomes important
again, at which point the instability restarts and drives turbulence
again (Sano \& Stone 2002b).

\section{Magnetically-driven Jets}

So far, I have focussed on the dynamics of an initially weak field
lying within the disk.  However, coupling of a large-scale poloidal
magnetic field to the rotation of the disk may produce a
centrifugally-driven outflow.  This mechanism has been suggested as
the origin of protostellar jets, though it is unclear whether the
outflow occurs just at the inner edge of the disk or extends over a
significant fraction of the disk (K\"onigl \& Pudritz 2000).  The
poloidal field could plausibly be the remains of the partial dragging
in of the original field threading the parent molecular cloud during
the core collapse and disk formation.

The centrifugal acceleration mechanism relies on coupling the magnetic
field to the disk and to the accelerated material, the field
transferring angular momentum from the former to the latter.  Clearly,
as with the magnetorotational instability, there must be sufficient
coupling for this mechanism to work.  In addition, there must be some
slippage so that the magnetic field is not dragged inwards by the
accreting material.  Thus there are constraints on the magnitude of
the diffusivity of the magnetic field.  What appears to be
unappreciated is that the \emph{nature} of the coupling (i.e.  whether
ambipolar, Hall or Ohmic diffusion is dominant) determines the
magnetic field direction as the lines emerge from the disk surface.
This in turn, controls the tendency of material to slide along the
field lines by the rotation (e.g.\ if the magnetic field lines are swept
back, the acceleration is less efficient).

Simulations are currently unable to solve the complete disk+protostellar jet
problem because of the large dynamic range in density between the disk
and wind.  Therefore disk-wind simulations are generally restricted to the
wind region, with the magnetic field, density and fluid velocity
specified at the base of the jet (roughly equivalent to the disk
surface).  Physically, these quantities are not independent but are
related through the diffusion of the field within the disk, which
simultaneously controls the bending of the field lines and how matter
is loaded onto them and lifted away from the disk (Wardle \& K\"onigl
1993; Wardle 1997).  This is still very much an open problem.

\section{Summary}

The examples discussed here -- magnetorotational instability and disk
winds -- show how the coupling between a magnetic field and the disk
material likely plays a crucial role in the structure and evolution of
protostellar disks.  It should be borne in mind that these two examples
need not be mutually-exclusive: within the disk the large-scale
poloidal field may be the mean of a field which is tangled (cf.\ Fig.\
3 of Blandford \& Payne 1982) by turbulence driven by the
magnetorotational instability.

The magnetic activity in protostellar disks is stratified because of
the strong dependence of diffusion on the level of ionisation.  This
implies that when grains are present, the bulk of the disk is
magnetically inactive within a few AU of the centre, where it is
self-shielded from ionisation sources.  Much of the magnetic activity,
and hence accretion, occurs in the surface layers of the disk (Gammie
1986).  When grains have settled to the midplane, Hall diffusion of
the magnetic field dominates over much of the disk (Balbus \& Terquem
2000, Sano \& Stone 2002b).  This tends to increase the extent of the
magnetically-coupled region within the disk, and reduces the extent of
the dead zone.

The separation of the disk into turbulent and inactive layers affects
planet-building, which is seeded by the coagulation and settling of
solid material to the midplane (e.g.\ Weidenschilling \& Cuzzi 1993).
This is complicated by the intriguing coupling between the grain
population and the magnetic field.  The grains determine the field
diffusivity and therefore affects the fluid dynamics, the fluid
motions in turn feed back on the grain population through the
competing processes of coagulation, shattering, mantle accumulation or
evaporation and settling mediated through the stirring and energy
dissipation by magnetically-driven turbulence in the disk.

The interdependence of the grains and fluid dynamics is, at present,
unexplored territory.  There is some hope for constraining theoretical
models using observational signatures of disk chemistry, which also be
dependent on the stratification of heating and the evolution of the
grain population.

\end{document}